\documentclass[prb,twocolumn,eqsecnum,showpacs]{revtex4}

\usepackage{graphicx}

\begin{document}

\title{Electromagnetic response and pseudo-zero-mode Landau levels \\
of bilayer graphene in a magnetic field}
\author{T. Misumi and K. Shizuya}
\affiliation{Yukawa Institute for Theoretical Physics\\
Kyoto University,~Kyoto 606-8502,~Japan }

\begin{abstract} 

The electromagnetic response of bilayer graphene in a magnetic field is studied
in comparison with that of monolayer graphene.
Both types of graphene turn out to be qualitatively quite similar in dielectric 
and screening characteristics, especially those deriving from vacuum fluctuations,
but the effect is generally much more sizable for bilayers.
The presence of the zero-(energy-)mode Landau levels is a feature specific to graphene. 
In bilayers, unlike in monolayers, the effect of the zero-mode levels
becomes visible and even dominant in density response 
as an externally-controllable band gap develops.
It is pointed out that the splitting of nearly-degenerate 
pseudo-zero-mode levels at each valley, specific to bilayer graphene, 
is controlled by an applied inplane electric field or by an injected current. 
In addition, a low-energy effective gauge theory of bilayer graphene 
is constructed.

\end{abstract}


\maketitle

\section{Introduction}

Graphene, a monolayer of graphite, has recently been attracting 
great attention, both experimentally~\cite{NG,ZTSK,ZJS} and
theoretically,~\cite{ZA,GS,PGN,NM,AF}
for its unusual electronic transport, characteristic of 
$\lq\lq$relativistic" charge carriers behaving like massless Dirac fermions. 
Graphene is naturally of interest from the viewpoint of relativistic
quantum field theory, and is a special laboratory 
to test the particle-hole picture~\cite{fnKlein} of the quantum vacuum
and, especially in a magnetic field, 
to study peculiar quantum phenomena~\cite{J,NS,Semenoff,Hal,FScs} 
tied to the chiral and parity anomalies.
Actually, the half-integer quantum Hall (QH) effect and 
the presence of the zero-energy Landau levels 
observed~\cite{NG,ZTSK} in graphene are a manifestation of
spectral asymmetry implied by the anomaly.

The bilayer (and  multilayer) of graphene is as interesting and 
exotic~\cite{MF,KA} as the monolayer. 
In bilayer graphene interlayer coupling modifies the intralayer relativistic spectra 
to yield  a quasiparticle spectrum with a parabolic energy dispersion.~\cite{MF}
The relativistic feature thereby disappears 
but the particle-hole structure still remains,
leading to a "chiral" Schroedinger Hamiltonian, 
which has no analog in particle physics.
Bilayer graphene, like monolayers, is intrinsically a gapless semiconductor
but has a notable property that the energy gap 
between the conduction and valence bands is 
controllable~\cite{OBSHR,Mc,CNMPL,MSBM,OHL} 
by use of external gates or chemical doping.

Theoretical studies show that electronic transport~\cite{ZA,GS,PGN}
and screening~\cite{Ando} in graphene are substantially different from
those in standard planar systems.
The difference becomes even more prominent under a magnetic field:~\cite{KSgr}
For graphene the vacuum state is a dielectric medium 
with appreciable electric and magnetic susceptibilities over all range of wavelengths.
Curiously the zero-energy Landau levels, 
though carrying normal Hall conductance $e^{2}/h$ per level, 
scarcely contribute to the dielectric effect.

The purpose of this paper is to study the electromagnetic response of 
bilayer graphene in a magnetic field at integer filling factor $\nu$, 
in comparison with that of monolayer graphene.
It turns out that both types of graphene are qualitatively quite similar in dielectric 
and screening characteristics, 
but the effect is much more sizable for bilayers 
(because of the difference in the basic cyclotron energy).

In bilayer graphene there arise two species of zero-energy levels 
(or zero modes), which, when an interlayer field is applied, move up or down 
(oppositely) at the two valleys.  
Remarkably, the dielectric effects due to the (pseudo) zero modes, unlike in monolayers, 
become visible and even grow steadily as the tunable band gap develops. 
It is pointed out that the splitting of nearly degenerate 
pseudo-zero-mode levels at each valley, specific to bilayer graphene, 
is controlled by an inplane electric field or by an injected current. 
In addition, we construct out of the response a low-energy effective gauge theory 
of bilayer graphene in a magnetic field and 
verify that the electric susceptibility of a QH system is generally expressed 
as a ratio of the Hall conductance to the Landau gap.

In Sec.~II we briefly review the low-energy effective theory 
of bilayer graphene and study its Landau-level spectrum.
In Sec.~III we examine the electromagnetic response and 
screening properties of bilayer graphene.
In Sec.~IV we derive an effective gauge theory.
In Sec.~V we study the effect of an inplane field on 
the almost degenerate zero-mode levels at each valley
and discuss its consequences.
Section~VI is devoted to a summary and discussion.

\section{Bilayer graphene}

The bilayer graphene consists of two coupled hexagonal lattices of carbon atoms,
arranged in Bernal $A'B$ stacking, with inequivalent sites denoted as 
$(A,B)$ in the bottom layer and $(A',B')$ in the top layer.
The electron fields in the bilayer are described by four-component spinors on the four sites 
and, as  in the case of monolayer graphene, their low-energy spectrum is governed 
by the electron states near the two inequivalent Fermi points $K$ and $K'$ 
in the Brillouin zone. 
The intralayer coupling $\gamma_{0} \equiv \gamma_{AB} \approx 2.9$ eV 
is related to the Fermi velocity in monolayer graphene, 
$v_{0} = (\sqrt{3}/2)\, a_{\rm L}\gamma_{0}/\hbar \approx 10^{6}$~m/s, 
with the lattice constant $a_{\rm L}=  0.246$nm.
The interlayer coupling 
$\gamma_{1} \equiv \gamma_{A'B}$ and
$\gamma_{3} \equiv \gamma_{AB'}$ are one-order of magnitude weaker
than the intralayer coupling $\gamma_{0}$; numerically,~\cite{MNEBP} 
$\gamma_{1} \approx 0.30$ eV and $\gamma_{3}\approx 0.10$ eV.

Actually the neighboring sites  $A'$ and $B$ form interlayer dimers via $\gamma_{1}$ 
and get shifted to higher energy bands, and the low-energy sector is 
essentially described by two-component spinors residing on the $A$ and $B'$ sites. 
The effective Hamiltonian is written as~\cite{MF}
\begin{eqnarray}
H&=&\!\! \int\! d^{2}{\bf x}\Big[ \psi^{\dag} ({\cal H}_{+}\! - eA_{0}) \psi 
+ \chi^{\dag} ({\cal H}_{-}\! - eA_{0}) \chi \Big], \nonumber\\
{\cal H}_{\xi} &=& {\cal H}_{0}  + {\cal H}_{\rm as}, \nonumber \\
{\cal H}_{0}&=&  \xi v_{3}\left(
\begin{array}{lc}
 & \Pi\\
 \Pi^{\dag}& \\
\end{array}
\right)
- {1\over{2m^{*}}}\left(
\begin{array}{lc}
 & (\Pi^{\dag})^{2}\\
 \Pi^{2}& \\
\end{array}
\right), \nonumber \\
{\cal H}_{\rm as}&=&\!\!  {\xi U\over{2}} \left[
\left( \! 
\begin{array}{ll}
1 & \\
 &\! -1\! \\ 
\end{array}
\right)
- {1\over{\gamma_{1}m^{*}}}\left( \! 
\begin{array}{ll}
\Pi^{\dag} \Pi & \\
 &\!\! -\Pi \Pi^{\dag} \!\!\\
\end{array}
\right) 
\right],
\label{Hbilayer}
\end{eqnarray}
where coupling to electromagnetic potentials 
$(A_{i}, A_{0})$ is introduced through 
$\Pi= \Pi_{x} -i \Pi_{y}$, $\Pi^{\dag}= \Pi_{x} + i\Pi_{y}$, and 
$\Pi_{i} = -i\partial_{i} + e A_{i}$.
Here $\xi=1$ refers to the $K$ valley with ${\cal H}_{+} ={\cal H}_{\xi=1}$
and $\psi = (\psi_{A}, \psi_{B'})^{\rm t}$
while $\xi=-1$ refers to the $K'$ valley with ${\cal H}_{-} ={\cal H}_{\xi=-1}$
and $\chi = (\chi_{B'}, \chi_{A})^{\rm t}$. We suppress the electron spin,
which is taken care of simply via spin degeneracy $g_{\rm s}=2$.

In ${\cal H}_{0}$ the first term with a linear dispersion represents 
direct interlayer hopping via $\gamma_{3}$, 
with characteristic velocity  
$v_{3} = (\sqrt{3}/2)\, a_{\rm L}\gamma_{3}/\hbar \sim  v_{0}/30$, 
while the second term with a quadratic dispersion represents 
$A \leftrightarrow B'$ hopping via the dimer state, giving 
$\hbar^{2}/(2m^{*}) = v_{0}^{2}/\gamma_{1}$.

The ${\cal H}_{\rm as}$ 
takes into account a possible asymmetry 
between the two layers, leading to a gap $U$ 
between the conduction and valence bands.
An important feature of bilayer graphene is 
that such a gap is controllable~\cite{OBSHR,Mc,CNMPL} 
by use of external gates, 
$U \approx e\triangle A_{0}$ with an interlayer 
voltage $\triangle A_{0}=A_{0}^{\rm top} -A_{0}^{\rm bottom}$.
The second term in ${\cal H}_{\rm as}$ is a layer asymmetry associated with 
the depleted charge on the $AB'$ dimer sites; we call it a kinetic asymmetry.

Let us place graphene in a strong uniform magnetic field $B>0$ 
normal to the sample plane; to this end we set 
$A_{i}(x)\rightarrow {\bf A}^{\! B}= B\, (-y,0)$.
 It is convenient to rescale $\Pi = \sqrt{2eB}\, a = (\sqrt{2}/\ell)\, a$ and 
$\Pi^{\dag} = (\sqrt{2}/\ell)\, a^{\dag}$ 
with the magnetic length $\ell= 1/\sqrt{eB}$ so that $[a, a^{\dag}] =1$. 
The kinetic terms thereby acquire the scale   
$v_{3} \rightarrow v_{3}\sqrt{2}/\ell \equiv \omega_{3}$ and 
$1/(2m^{*}) \rightarrow  eB/m^{*} \equiv \omega_{\rm c}$;
numerically, $\omega_{\rm c}\approx 3.9 \times B[{\rm T}]$ meV and 
$\omega_{3}\approx 1.2 \times \sqrt{B[{\rm T}]}$ meV,
where $B[{\rm T}]$ stands for the magnetic field in Tesla. 
The associated Landau-level spectra scale 
like $\omega_{3}\sqrt{n}$ and $\omega_{\rm c}\sqrt{n(n-1)}$ 
with the level index $n=0,1,\cdots$,  
with ratio $\sim 0.3/(\sqrt{n-1}\sqrt{B[{\rm T}]})$.
Therefore the $\omega_{3}$ term is practically negligible, 
compared with the $\omega_{\rm c}$ term,  
for higher Landau levels $n\ge 2$ under a strong magnetic field.
With this in mind we cast the Hamiltonian ${\cal H}_{\xi}$ in the form 
\begin{eqnarray}
{\cal H}_{\xi} &=&  \omega_{\rm c}\left(
\begin{array}{lc}
\mu\, (1 - z\, a^{\dag}a) &\lambda\, a -(a^{\dag})^{2}\\
 \lambda\, a^{\dag}-a^{2}& -\mu\, (1 -z\, aa^{\dag} ) \\
\end{array}
\right),
\end{eqnarray}
with $\lambda = \xi\,  \omega_{3}/\omega_{\rm c} 
\approx \pm 0.3 /\sqrt{B[{\rm T}]}$.
Here $\mu = \xi U/(2\omega_{\rm c})$ stands for half of the band gap 
in units of $\omega_{\rm c}$, and it seems feasible~\cite{Mc} to achieve 
an interlayer-voltage change of magnitude $\mu \sim O(1)$.
%
Note that the kinetic asymmtry  $\sim \mu\, z\, a^{\dag}a$ 
is very weak, with 
$z=2\omega_{\rm c}/\gamma_{1} \approx 0.026\times B[{\rm T}] \ll 1$.

Let us for the moment set a tiny parameter $z\rightarrow 0$. 
It is, then, generally seen from the structure of ${\cal H}_{\xi}$
that its eigenmodes are the same as those of ${\cal H}_{\xi}|_{\mu\rightarrow 0}$ 
and that the spectrum of ${\cal H}_{\xi}$ is symmetric about $\epsilon=0$, except for 
a possible $\epsilon= \omega_{\rm c} \mu$ spectrum or the zero-energy spectrum 
of  ${\cal H}_{\xi}|_{\mu\rightarrow 0}$.
We shall call the  $\epsilon=\pm \omega_{\rm c} |\mu|$ eigenmodes of ${\cal H}_{\xi}$
pseudo-zero-modes, or simply "zero" modes. 

Such spectra are slightly modulated by the tiny $O(\mu z)$ kinetic asymmetry.
For $\lambda=0$  it is possible to write down the eigenmodes explicitly.
The "nonzero" modes of ${\cal H}_{\xi}$ are Landau levels of energy 
\begin{eqnarray}
\epsilon_{n} =  s_{n}\, \omega_{\rm c} \sqrt{|n| (|n| -1) + \hat{\mu}_{n}^{2}}  
- {1\over{2}}\, \omega_{\rm c}\mu\, z
\label{specnonzero}
\end{eqnarray}
labeled by integers $n= \pm 2, \pm 3, \dots$, and
$p_{x}$ (or $y_{0} \equiv \ell^{2} p_{x}$); 
$\hat{\mu}_{n} = \mu \{ 1 - (|n| - {1\over{2}})\, z\}$.
Here $s_{n} \equiv {\rm sgn}\{n\} =\pm 1$ specifies 
the sign of the energy $\epsilon_{n}$. 
The associated eigenmodes are written as 
\begin{equation}
\psi_{n y_{0}}({\bf x}) = 
{1\over{\sqrt{2}}}\, \Big( c^{+}_{n} \phi_{|n|}({\bf x}), 
- s_{n}c^{-}_{n} \phi_{|n|-2}({\bf x})\Big)^{\rm t} ,
\end{equation}
where $c_{n}^{\pm} =\sqrt{1 \pm \hat{\mu}_{n}/\epsilon'_{n}}$
and $\epsilon'_{n} = s_{n}\, \sqrt{|n| (|n| -1) + \hat{\mu}_{n}^{2}}$; 
$\phi_{n}({\bf x})=\phi_{n}(y-y_{0})\, (e^{ix y_{0}/\ell^{2}}/\sqrt{2\pi \ell^{2}})$ 
are the eigenfunctions for the Landau levels $(n, y_{0})$ of the usual Hall electron.

In bilayer graphene there arise two nearly-degenerate zero-mode levels 
per valley and spin, with spectrum
\begin{eqnarray}
\epsilon_{0} &=& \omega_{\rm c}\mu,\  
\epsilon_{1} = \omega_{\rm c}\mu\, (1-z),  
\label{zeromodespectrum}
\end{eqnarray}
and eigenfunctions
$\psi_{0 y_{0}} =(\phi_{0}({\bf x}), 0)^{\rm t}$ 
and $\psi_{1y_{0}} = (\phi_{1}({\bf x}), 0)^{\rm t}$.
We take, without loss of generality,  
$U/\omega_{\rm c} = 2 \xi  \mu > 0$ and label the Landau levels associated 
with the $\xi =1$ valley (i.e., the $\psi$ sector with $\mu>0$) 
by $n=0_{+}, 1, \pm2, \cdots$
and those associated with the $\xi =-1$ valley (the $\chi$ sector
with $\mu<0$) by $n=0_{-}, -1, \pm2,  \cdots$.
Note that the zero-mode spectrum is ordered 
according to $\epsilon_{0_{-}} <\epsilon_{-1} < 0 < \epsilon_{1} <\epsilon_{ 0_{+}}$;
see Fig.~1.
An interlayer voltage $U \propto 2\mu$ thus works to shift the zero modes 
oppositely at the two valleys, opening a gap $U$,
while the nonzero-mode levels get shifted only slightly and remain 
symmetric, apart from the $O(\mu z)$ asymmetry.


\begin{figure}[tbp]
\begin{center}
\includegraphics[scale=0.52]{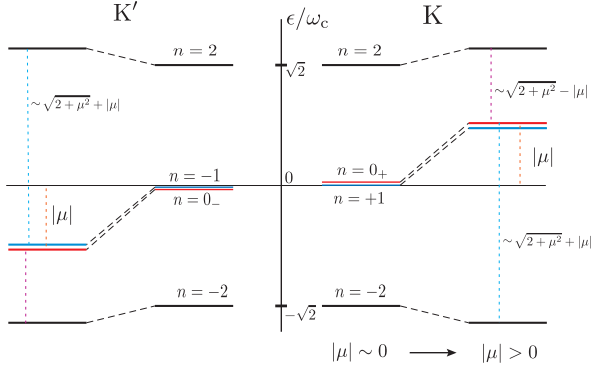}
\end{center}
\caption{(Color online)  Landau-level spectrum. The zero-mode levels move almost linearly 
with the interlayer voltage $\propto \mu$ and oppositely at the two valleys
while other levels are shifted only slightly.
}
\end{figure}


The zero-mode levels at each valley are degenerate for $z\rightarrow 0$ and 
the degeneracy remains even in the presence of the linear kinetic term.
This is because the zero modes persist even for $\lambda\not = 0$;
indeed, $D\phi=0$ with $D= \lambda a^{\dag} - a^{2}$ has two independent solutions 
of power series in $\lambda$,
\begin{eqnarray}
\phi_{0'}({\bf x}) &=& (1/N_{0})\sum_{n=0}^{\infty} 
\alpha_{n}\,  \phi_{3n}({\bf x}), \nonumber\\
\phi_{1'}({\bf x}) &=&  (1/N_{1}) \sum_{n=0}^{\infty} \zeta_{n}  \phi_{3n + 1}({\bf x}),
\nonumber\\
\alpha_{n}&=& \lambda^{n}\, \sqrt{{\Gamma(2/3)\, \Gamma(n+1/3)
\over{3^{n}\, n!\,  \Gamma (n+2/3)\, \Gamma (1/3)  }}},
\end{eqnarray}
with normalization factors $N_{0}^{2} =\sum_{n}\alpha_{n}^{2}$ 
and $N_{1}^{2} =\sum_{n}\zeta_{n}^{2}$;
$\zeta_{n}$ is given by $\alpha_{n}$ with replacement
$2/3\rightarrow 4/3$ and $1/3\rightarrow 2/3$ in the argument of the gamma functions.
These $n=0'_{+}, 1'$ solutions have an infinite radius of convergence in $\lambda$. 
This verifies that the index of the Dirac Hamiltonian,
${\rm Index}[{\cal H}_{\xi}|_{\mu\rightarrow 0}] = {\rm dim\ Ker} D 
- {\rm dim\  Ker} D^{\dag}$, is two, apart from the Landau-level degeneracy, 
\begin{equation}
{\rm Index}[{\cal H}_{\xi}|_{\mu\rightarrow 0}] = 2 \times {1\over{2\pi \ell^{2}}}
\int\! d^{2}{\bf x}
=\int\! d^{2}{\bf x}\,  {eB\over{\pi}}.
\label{indexth}
\end{equation}

The zero-mode spectrum, corrected by the $O(\mu z)$ asymmetry, eventually 
reads
\begin{eqnarray}
\epsilon_{0'} &=& \omega_{\rm c}\mu\, \{1- z\, b_{0}(\lambda)\}, \nonumber\\
\epsilon_{1'} &=& \omega_{\rm c}\mu\, \{1-z\, b_{1}(\lambda)\},
\end{eqnarray}
with $b_{0}(\lambda) = \sum_{n} 3 n\alpha_{n}^{2}/\sum_{n}\alpha_{n}^{2}$ and 
$b_{1}(\lambda) = \sum_{n}(3n+1)\zeta_{n}^{2}/\sum_{n}\zeta_{n}^{2}$, or
\begin{eqnarray}
b_{0}(\lambda) &=& (1/2)\, \lambda^{2} + 0.05\, \lambda^{4} + \cdots,
 \nonumber\\
b_{1}(\lambda) &=& 1+ (1/2)\, \lambda^{2} - 0.036\, \lambda^{4} + \cdots.
\label{bzerobone}
\end{eqnarray}
There is a common level shift of $O(z\lambda^{2})$
while the tiny level splitting $\sim \omega_{\rm c}\mu z \ll \omega_{\rm c}$
is practically unchanged (for $\lambda\sim 0.3$).
It is thus generally difficult to resolve the almost degenerate 
zero-mode levels; in a sense, it is the nonzero index~(\ref{indexth}) of 
the Hamiltonian ${\cal H}_{\xi}$ 
that underlies this stability in the zero-mode degeneracy.

In connection with the index, 
it would be worth remarking that $D^{\dag} \phi=0$ has no solution;  
a solution of infinite power series in $1/\lambda$, starting with $\phi_{0}({\bf x})$, 
fails to converge for finite $\lambda$.
Actually  the $\lambda\rightarrow \infty$ limit corresponds to the case 
of monolayer graphene with linear dispersion. 
The index thus jumps from "2" to "-1" as one passes from finite $\lambda$ 
to infinite $\lambda$.

The linear kinetic term affects the spectrum $\epsilon_{n}$ of the $|n| \ge 2$ levels
only to $O(\lambda^{2})$, which is still negligible in a strong magnetic field.  
We shall therefore set $\lambda\rightarrow 0$ in most of our analysis below, 
except in Sec.~V where we discuss possible resolution of the zero-mode levels.

For actual calculations it is useful to  make the Landau-level structure explicit
via the expansion~\cite{GJ,KSproj}
$\psi ({\bf x}, t) = \sum_{n, y_{0}} \langle {\bf x}| n, y_{0}\rangle\, 
\psi_{n}(y_{0},t)$. 
(From now on, we shall only display the $\psi$
sector since the $\chi$ sector is 
obtained by reversing the signs of $\mu$ and $\lambda$.)
The Hamiltonian $H$ thereby is rewritten as
\begin{eqnarray}
H\! &=& \!\! \int\! dy_{0} \!\!\!
\sum_{n =-\infty}^{\infty} \!\!\!
\psi^{\dag}_{n}\, \epsilon_{n}\, \psi_{n},
\label{Hzeronn}
\end{eqnarray}
and the charge density 
$\rho_{-{\bf p}}(t) =\int d^{2}{\bf x}\,  e^{i {\bf p\cdot x}}\,\psi^{\dag}\psi$ as 
\begin{equation}
\rho_{-{\bf p}} = e^{-\ell^{2} {\bf p}^{2}/4}
\sum_{k, n=-\infty}^{\infty} g_{k n}({\bf p})\int dy_{0}\,
\psi_{k}^{\dag}\, e^{i{\bf p\cdot r}}\,
\psi_{n} , 
\label{chargeoperator}
\end{equation}
with the coefficient matrix
\begin{eqnarray}
g_{k n}({\bf p}) &=& \textstyle{1\over{2}}\, \Big[
c_{k}^{+}\, c_{n}^{+}\, 
f_{|k|, |n|}({\bf p}) 
\nonumber\\   
&&\ \ \  
+ s_{k}s_{n}c_{k}^{-}\, c_{n}^{-}\, 
f_{|k| - 2,  |n| - 2}({\bf p}) \Big] ;
\label{gpsikn}
\end{eqnarray}
${\bf r} = (r_{1}, r_{2}) = (i\ell^{2}\partial/\partial y_{0}, y_{0})$
stands for the center coordinate with uncertainty 
$[r_{1}, r_{2}] =i\ell^{2}$.
Here 
\begin{equation}
f_{k n}({\bf p}) = \sqrt{{n!\over{k!}}}\, 
\Big({i\ell p\over{\sqrt{2}}}\Big)^{k-n}\, L^{(k-n)}_{n}
\Big(\textstyle{1\over{2}} \ell^{2}{\bf p}^{2}\Big)
\end{equation}
for $k \ge n$, and $f_{n k}({\bf p}) = [f_{k n}({\bf -p})]^{\dag}$;
$ p=p_{y}\! +\!ip_{x}$;
actually $f_{k n}({\bf p})$ are the coefficient functions 
for the charge density of the ordinary Hall electrons.

In a similar fashion one can derive the expression for the current operator
${\bf j}\sim \delta H/\delta {\bf A}$.
We omit it here, and simply remark that the current has 
a component coming from the $O(\mu z)$ asymmetry as well.

\section{Electromagnetic Response}

In this section we study the electromagnetic response of bilayer graphene.
Let us first consider the polarization function
$P({\bf p}, \omega) (\sim  -i\, \langle \rho \rho \rangle$) in Fourier space,
\begin{eqnarray}
P({\bf p}, \omega) 
&=& - \sum_{k,n}\Big\{ {1\over{\epsilon_{kn} -\omega}}
+ {1\over{\epsilon_{kn} +\omega}} \Big\}\,\sigma_{n}^{k}({\bf p}),
\nonumber \\ 
\sigma_{n}^{k}({\bf p})&=& {1\over{2\pi \ell^{2}}}\,
e^{- \ell^{2}{\bf p}^{2}/2} |g_{kn}({\bf p})|^{2},
\end{eqnarray}
where $\epsilon_{kn} = \epsilon_{k} - \epsilon_{n}$; 
the sum is taken over occupied levels $\{n\}$ and unoccupied levels $\{k\}$.
In what follows we focus on the real part of $P({\bf p}, \omega)$ 
in the static limit $\omega\rightarrow 0$, and denote the components 
coming from the virtual ($n\rightarrow k\rightarrow n$) transitions as
\begin{eqnarray}
P^{k}_{n}({\bf p})
&=& -{1\over{\pi \ell^{2}}}\,{1\over{\epsilon_{kn}}}\,
e^{- \ell^{2}{\bf p}^{2}/2} |g_{kn}({\bf p})|^{2}.
\label{Pkn}
\end{eqnarray}
Actually $|g_{kn}({\bf p})|^{2} = g_{nk}(-{\bf p})\, g_{kn}({\bf p})$
are functions of $\ell^{2}{\bf p}^{2}$ and are thus symmetric in $(k,n)$,
which implies the relation $P^{k}_{n} = -P^{n}_{k}$.

For conventional QH systems the polarization function vanishes for the vacuum 
since the charge operator trivially annihilates the vacuum $\rho |\nu=0\rangle =0$.
For graphene, in contrast, $\rho |\nu=0\rangle \not=0$ 
because of pair creation from the Dirac sea,
and even the vacuum acquires nonzero polarization
\begin{eqnarray}
P({\bf p}, 0)|_{\nu=0} 
&=& \sum_{k=0}^{N}\sum_{n=2}^{N} P^{k}_{-n}({\bf p}).
\label{vacuumPol}
\end{eqnarray}
Some care is needed in carrying out sums over an infinite number of Landau levels, 
which are potentially singular. 
For regularization we here choose, as before,~\cite{KSgr} 
to truncate the spectrum 
to a finite interval $-N\le n, k \le N$ and let $N\rightarrow \infty$ in the very end.
Regularization not only keeps calculations under control but also 
refines them:
Actually, instead of Eq.~(\ref{vacuumPol}), one may first consider,
for a given level $n$, the polarization "per level" 
$P_{n} = \sum_{k=-N}^{N} P_{n}^{k}$ by summing over all levels $k$ but $n$, 
and then  obtain $P({\bf p}, 0) = \sum_{n} P_{n}$ by summing over filled levels $n$.
This gives the same result as Eq.~(\ref{vacuumPol}) 
owing to the antisymmetry $P^{k}_{n} = -P^{n}_{k}$.

Equation~(\ref{vacuumPol}) is an expression for the $\psi$ sector 
($P \rightarrow P^{\psi}$). 
For the charge operator $\rho^{\chi} = \chi^{\dag}\chi$ 
in the $\chi$ sector one may replace, in Eq.~(\ref{chargeoperator}), 
$g_{kn}({\bf p})$ by 
\begin{eqnarray}
g^{\chi}_{kn}({\bf p}) = g_{-k,-n}({\bf p}),
\label{gxkn}
\end{eqnarray}
i.e., with the sign of $\mu$ reversed. 
One can define $(P^{\chi})^{k}_{n}({\bf p})$ accordingly and write 
the vacuum polarization function as 
$P^{\chi}({\bf p},0)|_{\nu=0} 
= \sum_{k=2}^{N}\sum_{n=0}^{N} (P^{\chi})^{k}_{-n}({\bf p})$.
Note that these components enjoy the following property
\begin{equation}
(P^{\psi})_{-n}^{k} = - (P^{\psi})_{k}^{-n} = (P^{\chi})_{-k}^{n}.
\end{equation}
This implies, in particular, that the $\psi$ and $\chi$ sectors contribute equally 
to the vacuum polarization $P^{\psi}({\bf p}, 0)|_{\nu=0} = P^{\chi}({\bf p}, 0)|_{\nu=0}$, 
and also that $P({\bf p}, 0) = P^{\psi}({\bf p}) + P^{\chi}({\bf p})$ 
is the same for the charge-conjugate states with $\nu = \pm$ integer. 
(In view of this, we shall focus on the case $\nu\ge0$ from now on.)

In calculating the density response we suppose that 
the almost degenerate zero-mode levels at each valley 
remain practically inseparable and treat them 
as both occupied or empty; accordingly we set $z\rightarrow 0$ below.
Let us first look into the leading long-wavelength part $\sim  O({\bf p}^{2})$ of 
$P({\bf p}, 0)$, to be denoted as $P^{(2)}({\bf p}, 0)$, 
which is related to the electric susceptibility 
$\alpha_{\rm e} = - (e^{2}/{\bf p}^{2})\, P^{(2)}({\bf p},0)$.
A look into the matrix elements in Eq.~(\ref{gpsikn}) shows 
that $P^{(2)}_{n}({\bf p}, 0)$ derives only 
from virtual transitions to the adjacent levels 
($n \rightarrow n\pm 1$) and the related ones across the Dirac sea
$(n\rightarrow  - n \pm 1)$.
A direct calculation then yields
\begin{eqnarray}
P^{(2)}_{n}
 &=& -{{\bf p}^{2}\over{2\pi\, \omega_{\rm c}}}\,\Big( \beta_{n} 
- {\mu\over{|n|}(|n| -1)} \Big),   
\label{Pntwo}\\
\beta_{n} &=& s_{n} {|n|-1/2\over{\sqrt{|n| (|n|-1) + \mu^{2}}} } , 
\end{eqnarray}
for $2\le |n|\le N-1$, and   
\begin{equation}
P^{(2)}_{1} + P^{(2)}_{0_{+}} =
 -({\bf p}^{2}/2\pi\omega_{\rm c})\, 2 |\mu| .
\label{Pzerotwo}
\end{equation}
The bottom of the Dirac sea yields 
$P^{(2)}_{-N} = (P^{(2)})_{-N}^{N-1} 
+(P^{(2)})_{-N}^{ -(N-1)} \propto N-1 + O(1/N)$, 
which properly makes 
$P^{(2)}|_{\nu=0} = \sum_{n=2}^{N} P^{(2)}_{-n}$ finite.
This leads to the vacuum electric susceptibility
(per valley and spin)
\begin{eqnarray}
\alpha_{\rm e}^{\rm vac} 
 &=&  {e^{2} \over{2\pi\, \omega_{\rm c}}}\, F(\mu), \\
F(\mu) &=& -\sum_{n=2}^{N-1} \beta_{n} - |\mu| + N-1,   \nonumber\\
&\approx& 0.87715
 - |\mu| + O(\mu^{2}).
\end{eqnarray}

An alternative expression for this $\alpha_{\rm e}^{\rm vac}$ is obtained 
by summing up the virtual
$(-n \rightarrow  n \pm 1)$ processes, or
$\sum_{n=2}^{N} P_{-n}^{n-1} + \sum_{n=2}^{N-1}P_{-n}^{n+1}$, 
which yields 
\begin{eqnarray}
F(\mu)\!\! &=&\!\!
\sum_{n=1}^{N-1} {1\over{4\, \epsilon'_{n}\epsilon'_{n+1}}} \Big[
2 (\epsilon'_{n+1} - \epsilon'_{n}) - (\epsilon'_{n+1} - \epsilon'_{n})^3  \Big]
\nonumber\\
&=&\!\! \sum_{n=1}^{N-1} 
{(\sqrt{n+1} -\sqrt{n\!-\!1})^{3}\over{4\sqrt{n}}} - |\mu|  + O(\mu^{2}),
\end{eqnarray}
where $\epsilon'_{n} = \sqrt{n (n -1) + \mu^{2}}$.

Equations~(\ref{Pntwo}) and~(\ref{Pzerotwo}) tell us that 
the susceptibility carried by a positive-energy level of the same $n$ is
slightly different by $O(\mu)$ terms at the two valleys.
Such $O(\mu)$ corrections are visible only for the zero-mode levels 
since the two valleys are practically indistinguishable for $n\ge 2$ 
(up to $O(\mu z)$ splitting).  
Let us suppose that the electrons fill up an integral number $\nu$
of Landau levels, with uniform density
$\langle \rho\rangle \equiv \bar{\rho} = \nu /(2\pi \ell^{2})$.
We write $\nu = \sum_{n} \nu_{n}$ in terms of the filling factors
$\nu_{n}$ of the $n$th level $[\, 0\le \nu_{n} \le 4$ for $n\ge 2$ and 
$0\le \nu_{\{1,0\}} \le 4; \nu_{\{1,0\}}\equiv \nu_{1} + \nu_{0_{+}}]$,
with both valley and spin taken into account. 
The electric susceptibility at integer filling $\nu$ is then written as
\begin{equation}
\alpha_{\rm e}
 =  {e^{2} \over{2\pi\, \omega_{\rm c}}}\,
\Big\{ 4 F(\mu)  + \nu_{\{1,0\}} \, |\mu| 
+ \sum_{n\ge 2} \nu_{n}\, \beta_{n} 
\Big\}.
\label{alphae}
\end{equation}
(This $\alpha_{\rm e}$ is even in $\nu$,
and applies to the case of holes with $\nu\le 0$ equally well.)

It is amusing to note here that some expressions greatly simplify 
for a special value $\mu = 1/2$: The spectrum is equally spaced, 
$\epsilon_{n} = s_{n} \omega_{\rm c} (|n| - 1/2)$ for $|n| \ge 2$ 
and $\epsilon_{1} = \epsilon_{0_{+}} = \omega_{\rm c}/2$,
apart from the $O(\mu z)$ splitting.  For the susceptibility 
one finds $\beta_{n}= 1$ for $n\ge 2$, and $F(1/2) = 1/2$.
This yields
\begin{equation}
\alpha_{\rm e}|_{\mu=1/2} =  
(2 \delta_{\nu\, 0} + \nu)\, e^{2}/(2\pi\, \omega_{\rm c}),
\end{equation}
which rises linearly with $\nu=4,8,12, \cdots$.


\begin{figure}[tbp]
\begin{center}
\includegraphics[scale=1]{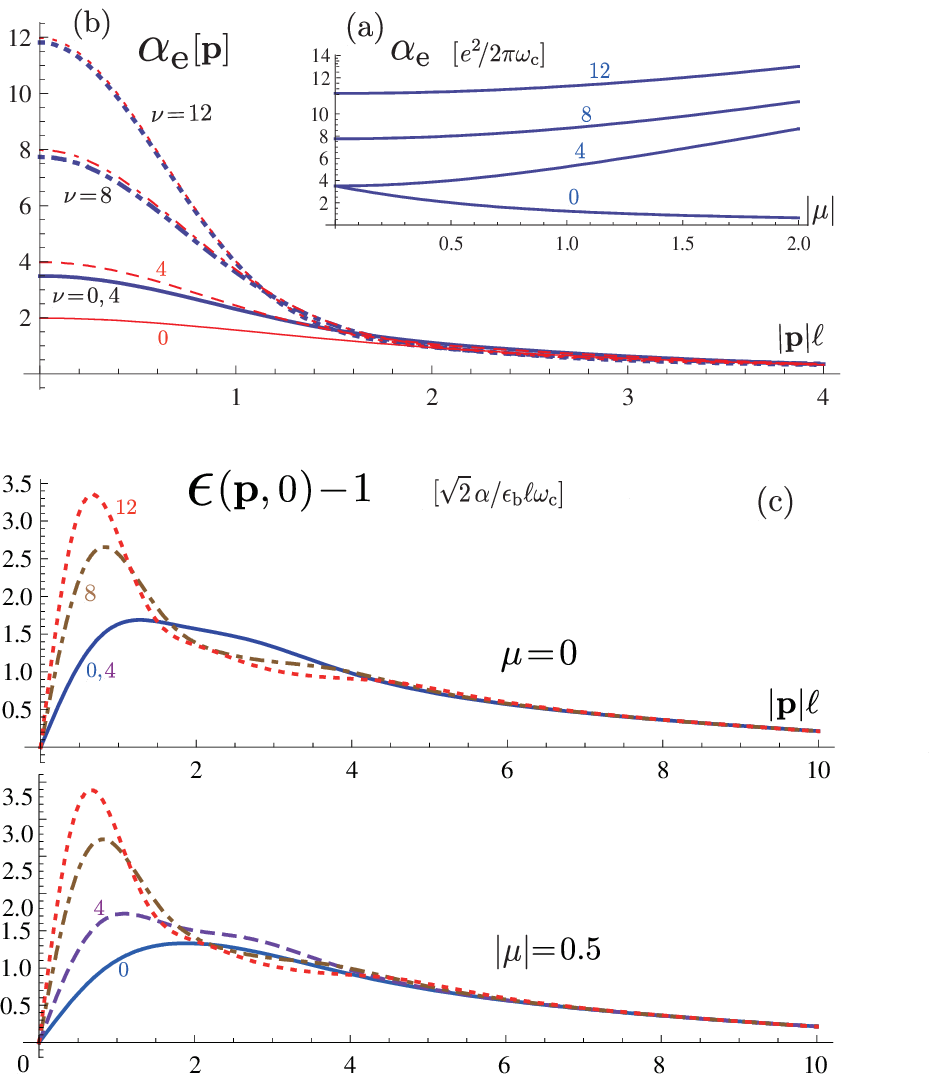}
\end{center}
\caption{(Color online) (a) Electric susceptibility $\alpha_{\rm e}$ 
decreases with the band gap $2|\mu|\, \omega_{\rm c}$ for $\nu=0$ and 
rises for $\nu=$4, 8 and 12.
(b) Static susceptibility function $\alpha_{\rm e} [{\bf p}]$, 
in units of $e^{2}/(2\pi\, \omega_{\rm c})$, 
at $\nu= 0, 4, 8, 12$ for zero band gap $\propto \mu=0$ (thick curves) and 
for a finite gap $|\mu|=0.5$ (thin red curves). 
(c) Static dielectric function $\epsilon ({\bf p},0) -1$,
in units of $\sqrt{2}\alpha/ (\epsilon_{\rm b}\omega_{\rm c} \ell)$, 
at $\nu= 0 \sim 12$ for a band gap  
$\propto \mu=$ 0 and $|\mu| =0.5$.
}
\end{figure}


In Fig.~2~(a) we plot $\alpha_{\rm e}$ as a function of 
the band gap $2 |\mu|\, \omega_{\rm c}$ for $\nu=0, 4, 8$ and 12. 
The vacuum susceptibility $\alpha^{\rm vac}_{\rm e} =\alpha_{\rm e}|_{\nu=0}$ is
almost comparable to the contribution of the filled $n=2$ level for $\mu\sim 0$.
The $\alpha^{\rm vac}_{\rm e}$ decreases gradually 
as the band gap $\propto \mu$ develops; 
this is clear intuitively.
In contrast, when the zero-mode levels get filled, i.e., at $\nu=4$,  
$\alpha_{\rm e}$ starts to grow almost linearly with $\mu$, 
and such characteristic behavior persists for higher $\nu$ as well.
This is somewhat unexpected but is easy to understand: 
The zero-mode levels move linearly with the band gap $\propto |\mu|$ while the $n\ge 2$ 
levels are shifted only slowly, as seen from Fig.~1.
The growth of $\alpha_{\rm e}$ is therefore due to 
a decrease in activation gap from the zero-mode levels.  
This implies that the dielectric effect due to the zero modes, 
though negligible for $|\mu| \ll 1$, becomes dominant for large gaps $|\mu| \sim 1$.

This is also the case with the full expression for the polarization 
$P({\bf p}, 0)$,
which one can evaluate numerically using Eqs.~(\ref{gpsikn}) and (\ref{Pkn}).
In Fig.~2 (b) we plot the susceptibility function 
$\alpha_{\rm e}[{\bf p}]  = - (e^{2}/{\bf p}^{2})\, P({\bf p},0)$ 
at $\nu=$0, 4, 8, 12 for zero band gap and for a finite gap.
There, as in the monolayer case, the zero-mode levels scarcely contribute 
to $\alpha_{\rm e}[{\bf p}]$ for zero gap, but 
one sees clearly that,   as the band gap $\propto \mu$ develops, 
they make $\alpha_{\rm e}[{\bf p}]$ distinct
between the vacuum and the $\nu=4$ state.

The susceptibility is also  related to screening properties of graphene.  
Let us now turn on the Coulomb interaction $v=\alpha/(\epsilon_{\rm b}|{\bf x}|)$ 
or $v_{\bf p} = 2\pi \alpha/(\epsilon_{\rm b} |{\bf p}|)$ 
with $\alpha = e^{2}/(4\pi \epsilon_{0}) \approx 1/137$ 
and the substrate dielectric constant $\epsilon_{\rm b}$, and 
study its effects in the random-phase approximation (RPA).  
The RPA dielectric function is written as~\cite{mahan} 
\begin{equation}
\epsilon ({\bf p}, \omega) 
= 1 - v_{\bf p}\, P({\bf p}, \omega).
\end{equation}
Figure~2 (c) shows the static function 
$\epsilon({\bf p},0) -1$ for $\nu=$ 0, 4, 8, 12,
plotted in units of $\sqrt{2}\alpha/ (\epsilon_{\rm b}\omega_{\rm c} \ell)$.
Note first that there is no screening at long distances, 
$\epsilon({\bf p},0)  \rightarrow 1$ for ${\bf p}\rightarrow 0$, 
as is typical of two-dimensional systems.
As wave vector $|{\bf p}|$ is increased, 
$\epsilon({\bf p},0)$ grows rapidly, becomes sizable
for $|{\bf p}|\ell \sim 1$, and then decreases only gradually 
for larger $|{\bf p}|$.
Such profiles of $\epsilon ({\bf p},0)$ 
and $\alpha_{\rm e}[{\bf p}]$ in Fig.~2  
are qualitatively quite similar to those of monolayer graphene 
studied earlier.~\cite{KSgr}

Still there are some clear differences:
(1) Note that the basic Landau gap of bilayer graphene,
$\omega_{\rm c}\approx 45\, B[{\rm T}]$ K,
is about one-order of magnitude smaller than 
the monolayer gap   
$\omega_{\rm c}^{\rm mono} \approx 400\,\sqrt{ B[{\rm T}]}$ K at $B$=1T, or
\begin{equation}
\omega_{\rm c}/\omega_{\rm c}^{\rm mono} 
\approx 0.1 \sqrt{B[{\rm T}]}.
\label{GaAsvsgraph}
\end{equation}
Numerically it turns out that, for both monolayer and bilayer graphene, 
the vacuum susceptibility $\alpha_{\rm e}^{\rm vac}$ is around 3 
in units of $e^{2}/(2\pi\, \omega_{\rm c})$  and the  peak value of 
$\epsilon ({\bf p},0)|_{\nu=0}-1$ is around 
1.8 in units of $\sqrt{2}\alpha/ (\epsilon_{\rm b}\omega_{\rm c} \ell)$. 
This actually means that $\epsilon ({\bf p},0)$ 
and $\alpha_{\rm e}[{\bf p}]$ are numerically more sizable for bilayers 
than for monolayers.
In particular, for the bilayer the peak value of $\epsilon ({\bf p},0)$ 
in Fig.~2 (c) would range from 
$\epsilon({\bf p},0)\approx 9.5
\sim 18$ for $\nu = 0 \sim 12$, with the choice 
$\epsilon_{\rm b}\approx 4$ and $B=$ 1T, or 
$\sqrt{2}\alpha/ (\epsilon_{\rm b}\omega_{\rm c} \ell) \approx 5.1$.
(The value of the substrate dielectric constant $\epsilon_{\rm b}$ 
depends on the structure of the sample;  
with SiO$_{2}$ on both sides of the bilayer, e.g., one can set
$\epsilon_{\rm b}\approx \epsilon_{{\rm SiO}_{2}}\approx 4$.)
This implies that the Coulomb interaction is very efficiently screened 
(and weakened) in bilayer graphene.

(2) For graphene, unlike standard QH systems, even the  $\nu=0$ vacuum state
has an appreciable amount of polarization $\epsilon({\bf p},0) -1$ 
over a wide range of wavelengths, which reflects the quantum fluctuations 
or "echoes" of the Dirac sea in response to an applied field. 
The echoes are in a sense "harder" for monolayer graphene~\cite{KSgr}
for which $\epsilon({\bf p},0)$ appears almost constant over the wave-vector range 
in Fig.~2 (c).
The rise of the peak values of $\epsilon({\bf p},0) -1$ 
with filling factor $\nu$ is more prominent for monolayers than bilayers;
see Eq.~(\ref{aevsSigmaHall}).
These features reflect the difference in the underlying Landau-level structures.

(3) In  bilayer graphene, unlike in monolayers, 
the effects of the zero-mode Landau levels 
become visible as the band gap $\propto 2 |\mu|$ develops and this makes 
the $\nu=0$ and $\nu=4$ states distinguishable in $\epsilon ({\bf p},0)$ 
and $\alpha_{\rm e}[{\bf p}]$.


\begin{figure}[tbp]
\begin{center}
\includegraphics[scale=0.6]{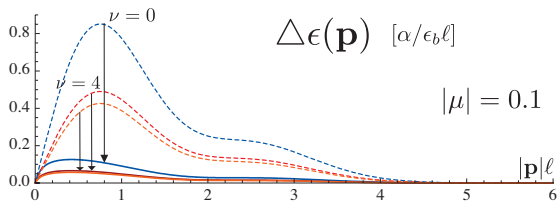}
\end{center}
\caption{(Color online) Exciton spectra at $\nu$=0 and 4 for $|\mu|=0.1$,
in units of $\alpha/\epsilon_{\rm b}\ell$. 
The spectra are reduced from the bare ones (dashed curves) 
via screening.
At $\nu=4$ there are two spectra with excitation gaps 
$\sim\omega_{\rm c}(\sqrt{2+\mu^{2}} \mp |\mu|)$.}
\end{figure}


In the RPA the response function is written as $P_{\rm RPA}({\bf p},\omega) 
=P({\bf p},\omega)/\epsilon ({\bf p}, \omega)$, from which   
one can derive~\cite{KH} the inter-Landau-level excitation spectra 
corrected by the Coulomb interaction.
Isolating from $P({\bf p}, \omega)$ one of its poles at
$\omega \sim \epsilon_{k} -\epsilon_{n}$ 
and setting $\epsilon ({\bf p}, \omega)  = 1 - v_{\bf p}\, P({\bf p},
\omega) \rightarrow 0$ fixes the pole position of $P_{\rm RPA}({\bf p},\omega)$, 
$\epsilon_{k,n}^{\rm RPA} = \epsilon_{k} -\epsilon_{n} + \triangle
\epsilon_{k,n}({\bf p})$ with 
\begin{eqnarray}
\triangle \epsilon_{k,n}({\bf p})
&\approx& {\alpha\over{\epsilon_{\rm b}\ell}}\, 
{ \ell |{\bf p}|\over{2\, \epsilon ({\bf p},0)}}\, \nu_{\rm g}({\bf p}), 
\nonumber\\
\nu_{\rm g}({\bf p})\! &=&\!\!\! \sum g_{nk}(-{\bf p})
g_{kn}({\bf p})\, e^{-x}/x,
\label{excitonspec}
\end{eqnarray}
where $x= \ell^{2}{\bf p}^{2}/2$.

The vacuum state at $\nu=0$ supports excitons 
associated with the $n=-2 \rightarrow (1, 0_{+})$ or 
$(-1, 0_{-})\rightarrow 2$ transitions with the excitation gap 
$\epsilon \approx \omega_{\rm c} (\sqrt{2 + \mu^{2} } + \mu)$ 
and 
\begin{equation}
\nu_{\rm g}({\bf p}) = 4\, (1- \mu /\sqrt{2 + \mu^{2}}\, )\,  h(x) ,
\label{nugp}
\end{equation}
where $ h(x) = e^{-x}\,  \{1+ {1\over{4}}\,  x(x-3) \}$.
The $\nu=4$ state supports excitons with energy 
$\epsilon \approx \omega_{\rm c} (\sqrt{2 + \mu^{2} } \pm \mu)$ and 
$\nu_{\rm g}({\bf p}) = 2\, (1\mp \mu /\sqrt{2 + \mu^{2}}\, )\,  h(x)$,
respectively; see Fig.~1.
Similarly, at $\nu=4n$ ($n\ge 2)$ the $n\rightarrow n+1$ transition 
with a gap $\triangle\omega_{\rm c}^{(n)} 
\approx \epsilon_{n+1} - \epsilon_{n}$ 
gives rise to excitons with  
$\nu_{\rm g}(0) = (\sqrt{n-1} +\sqrt{n+1})^{2}$.
As a result of screening, 
$\triangle \epsilon_{k,n}({\bf p})\propto 1/\epsilon ({\bf p},0)$,
one would observe prominent reduction in magnitude of 
the exciton spectra; see Fig.~3.

\section{Effective gauge theory}

In this section we study low-energy response of bilayer graphene
and construct an effective gauge theory. 
Let us consider the Hall conductance which is read 
from a response of the form 
${1\over{2}}\sigma_{\rm H} (A_{x}\partial_{t}A_{y}- A_{y}\partial_{t}A_{x})$;
it is calculated from the current-current correlation function 
and from Berry's phase as well.~\cite{FScs}
A direct calculation similar to the one in the monolayer case~\cite{KSgr} 
shows that $\sigma_{\rm H} \rightarrow e^{2}\ell^{2}$ per electron so that 
$\sigma_{\rm H} \rightarrow e^{2}/(2\pi \hbar)=e^{2}/h$ per filled level.
The result is independent of $\lambda$ and $z$. 
 
Care is needed to determine $\sigma_{\rm H}$ carried by the vacuum state.  
As before, we truncate the spectrum and find that the bottom of 
the Dirac sea contributes $\sim -(N-1) e^{2}/h$.
This yields the vacuum Hall conductance
\begin{equation}
\sigma_{\rm H}^{\rm vac}(\xi = \pm1) = \mp e^{2}/h
\end{equation}
per valley and spin. This implies that nonzero current and charge are 
induced~\cite{Semenoff,Hal, FScs}in the vacua at the two valleys, but they combine 
to vanish in the vacuum, leaving no observable effect.
In general, the Hall conductance $\sigma_{\rm H}$ is cast in the form of 
a spectral asymmetry and the vacuum Hall conductance 
$\sigma_{\rm H}^{\rm vac}$ is related to half of the index~(\ref{indexth}).

The long-wavelength response of bilayer graphene 
with uniform electron density
$\bar{\rho}$ is now summarized
by the Lagrangian
\begin{eqnarray}
L_{A} &=&  \bar{\rho}\, e A_{0}
- e^{2}\ell^{2}\bar{\rho}\, {1\over{2}}\,
\epsilon^{\mu\nu\rho}\, A_{\mu}\partial_{\nu}A_{\rho}
\nonumber\\
&& + {1\over{2}}\,
\alpha_{\rm e}\,  {\bf E}_{\parallel}^{2} -{1\over{2}}\,\alpha_{\rm
m}\, (A_{12})^{2}.
\label{emresponse}
\end{eqnarray}
Here $A_{0}$ detects the charge density $\bar{\rho}$, 
$\alpha_{\rm e}$ is the electric susceptibility in Eq.~(\ref{alphae}) 
and $\alpha_{\rm m} \propto  e^{2}\ell^{2}\bar{\rho}/m^{*}$ stands for 
the magnetic susceptibility probed by a local variation 
$A_{12}= \partial_{x}A_{y} - \partial_{y}A_{x}$ about $B$.

This response is essentially the same as that for monolayer graphene,~\cite{KSgr} 
except for the values of  $\alpha_{\rm e}$ and $\alpha_{\rm m}$.
Accordingly the effective theory also takes the same form,
i.e., a theory of a vector field $b_{\mu}=(b_{0}, b_{1}, b_{2})$, 
with the Lagrangian to $O(\partial^{2})$,
\begin{eqnarray}
L_{\rm eff}[b] 
&=& -e A_{\mu}\, \epsilon^{\mu\nu\lambda}
\partial_{\nu} b_{\lambda} + {1\over{\ell^{2}}}\, b_{0}
+ {1\over{2\ell^{2}\bar{\rho}}}\, b_{\mu}\, \epsilon^{\mu\nu\lambda}
\partial_{\nu} b_{\lambda}  \nonumber\\
&&
+ {1\over{2\ell^{2}\bar{\rho}\, \omega_{\rm eff}}}\, 
(b_{k0})^{2} - {1\over{2}}\, \delta b_{12}\, v \, \delta b_{12}
\label{efftheory}
\end{eqnarray}
and the {\sl effective} cyclotron frequency 
\begin{eqnarray}
\omega_{\rm eff} &=& e^{2}\ell^{2}\bar{\rho} /\alpha_{\rm e}
=\omega_{\rm c}\, g(\nu),  \nonumber\\
g(\nu) &=&
{\nu \over{  4 F(\mu^{2})  + \nu_{\{1,0\}}\, |\mu| 
+ \sum_{n\ge 2} \nu_{n}\, \beta_{n}} } ,
\label{weff}
\end{eqnarray}
where $b_{\mu\nu} = \partial_{\mu}b_{\nu} -
\partial_{\nu}b_{\mu}$;
$\ell^{2}\bar{\rho} =  \nu/(2\pi)$ and $\nu= \sum
\nu_{n}$.
An advantage of bosonization~\cite{funcBos,LZ} is that 
it allows one to handle the Coulomb interaction exactly;
it is included in Eq.~(\ref{efftheory}) with shorthand notation
$\delta b_{12}\, v \, \delta b_{12} = \int d^{2}{\bf y}\,
\delta b_{12}(x)\, v({\bf x\! -\! y}) \delta b_{12}(y)$ and
$\delta b_{12} = b_{12} - \bar{\rho}$.
(We have omitted the $\alpha_{\rm m}$ term from $L_{\rm eff}[b]$  
since the Coulomb interaction overtakes it at long wavelengths.)

This effective Lagrangian not only 
reproduces the original response~(\ref{emresponse}),
but also shows that the Coulomb interaction 
$v_{\bf p} = 2\pi \alpha/(\epsilon_{\rm b} |{\bf p}|)$ 
substantially modifies the dispersion of 
the cyclotron mode at long wavelengths ${\bf p} \rightarrow 0$,
\begin{equation}
\omega ({\bf p})
\approx  \omega_{\rm eff} 
+ {1\over{2}}\, (\ell^{2}\bar{\rho}\, v_{\bf p} 
+\cdots )\, {\bf p}^{2},
\label{Coulombshift}
\end{equation}
where $\cdots$ involves $\alpha_{\rm m}/(\alpha_{\rm e} \omega_{\rm eff})$, 
if $\alpha_{\rm m}$ is recovered.
This excitation spectrum is in good agreement with 
the RPA result~(\ref{excitonspec}) at long wavelengths;
the filling factor $\nu=2\pi \ell^{2}\bar{\rho}$ in Eq.~(\ref{Coulombshift}) 
corresponds to $\nu_{\rm g}(0)$ in Eq.~(\ref{excitonspec}).

For graphene the Landau levels are not equally spaced and the excitation
gaps depend on the level index $n$ or $\nu$.
At $\nu= 4n\  (n\ge 1)$, e.g., the minimum gap is
$\triangle\omega_{\rm c}^{(n)} = (\sqrt{n(n+1) + \mu^{2}} 
-\sqrt{n(n-1)+ \mu^{2} } )\, \omega_{\rm c}$, and 
$\omega_{\rm eff}$ in Eq.~(\ref{weff}) represents such an activation gap.

In particular, at $\nu=4$ where the zero-mode levels are filled,
the effective gap $\propto g(4) = 1/[F(\mu) +\mu]$ derives solely from
the vacuum fluctuations; it deviates from the true gap
$\triangle\omega^{(1)}_{\rm c} 
= (\sqrt{2+ \mu^{2}} -\mu)\, \omega_{\rm c}$ 
by about 20\% for $\mu=0$ and by  5\% or less for $\mu>0.4$.
At $\nu=8$, $\omega_{\rm eff}$ and $\triangle\omega_{\rm c}^{(2)}$
agree within 1.1\% for all $\mu$. The agreement is almost exact for higher $\nu$.
It is somewhat surprising that, as in the monolayer case, an effective theory constructed 
from the long-wavelength response alone gives an excellent description of 
the excitation spectrum.
Here we verify again from Eq.~(\ref{weff}) that 
the susceptibility $\alpha_{\rm e}$ of a QH system is generally
given by a ratio of the Hall conductance to the Landau gap,~\cite{KSgr}
\begin{equation}
\alpha_{\rm e} = \nu (e^{2}/h) /\omega_{\rm eff} \approx \sigma_{\rm H}/
\triangle\omega_{\rm c}^{(n)}.
\label{aevsSigmaHall}
\end{equation}
This implies that $\alpha_{\rm e}$ depends on $\nu$ and $\mu$
nontrivially while it grows like $\alpha_{\rm e} \propto \nu^{3/2}$ 
for monolayer graphene.

\section{Zero-mode Landau levels}

In this section we take a close look into the properties of
the zero-mode levels at each valley, with a tiny splitting 
$\omega_{\rm c}\mu z \ll \omega_{\rm c}$.
Special care is needed in considering the response from such an almost degenerate sector,
since even a weak field applied as a probe may affect the true eigenmodes.
The first task therefore is to resolve the degeneracy of the $(0'_{+},1')$ sector 
by diagonalizing  the external probe plus the $O(\mu z)$ asymmetry,
\begin{equation}
\delta H =-\int\! d^{2}{\bf x}\,  e A_{0} \rho + 
\int\! dy_{0}\,\psi^{\dag}\triangle{\cal H}_{\rm as}\psi,
\end{equation}
where $\triangle{\cal H}_{\rm as} =\omega_{\rm c}\mu z\,{\rm diag}[- a^{\dag}a, a a^{\dag}]$.
We suppose that $A_{0}$ is slowly varying in space  and retain only terms up to first order in 
$\partial_{{\bf x}}$, i.e.,  ${\bf E} =-\partial_{\bf x}A_{0}$.  
In practice, it is convenient, without loss of generality, 
to take $A_{0}({\bf x},t)\rightarrow A_{0}(y,t)$, or 
${\bf E} \parallel E_{y}$.

We take the linear kinetic term $\propto \lambda$ into account 
(since it may potentially be important at such a low energy scale).
Within the $(0'_{+},1')$ sector $\delta H$ effectively turns~\cite{fnone} 
into the matrix Hamiltonian (in $y_{0}$ space) 
\begin{equation}
- e A_{0}(y_{0},t)  - \kappa \, b_{+}(\lambda) + \left(
\matrix{\kappa\, b_{-}(\lambda) & {\cal E}\, c(\lambda) \cr  {\cal E}\, c(\lambda)
& -\kappa\,b_{-}(\lambda)}
\right) , 
\label{Hdeg}
\end{equation}
with ${\cal E}\equiv e \ell E_{y}/\sqrt{2}$, $\kappa = (1/2)\, \omega_{\rm c} \mu\, z$;
$c(\lambda) = \sum_{n=0}^{\infty} \sqrt{3n +1}\alpha_{n} \zeta_{n}/
(N_{0}N_{1})$.
Here $b_{\pm}(\lambda) = b_{1}(\lambda) \pm  b_{0}(\lambda)$ in terms of 
$b_{0}$ and $b_{1}$ defined in Eq.~(\ref{bzerobone}).
Numerically, 
\begin{eqnarray}
c(\lambda) &=& 1- 3 \times 10^{-5} \lambda^{4} 
+\cdots, \nonumber\\
b_{+}(\lambda) &=& 1 + \lambda^{2} + 0.086\, \lambda^{4}
+\cdots, \nonumber\\
b_{-}(\lambda) &=&  1- 0.014\, \lambda^{4} + \cdots.
\end{eqnarray}
The  $O(\lambda^{4})$ corrections are practically negligible 
for $\lambda\sim 0.3$.
We therefore set $c(\lambda) \approx b_{-}(\lambda) \approx 1$ and 
may keep the $O(\lambda^{2})$ correction to $b_{+}(\lambda)$.

The Hamiltonian~(\ref{Hdeg})   
leads to the level splitting $\pm \sqrt{ \kappa^{2} + {\cal E}^{2}}$
with the new eigenmodes $0_{+}''$ and $1''$  related to $0_{+}'$ and $1'$ 
via the unitary transformation
$\psi_{0''_{+}} = \psi_{0'_{+}} \cos \theta - \psi_{1'}\sin \theta$,
with $\tan2 \theta = e\ell\, E_{y}/(\sqrt{2}\, \kappa)$.
Here we see that an inplane electric field enhances the splitting of 
the zero-mode levels, with a gap  
\begin{equation}
\triangle \epsilon  = \omega_{\rm c}\sqrt{(\mu\, z)^{2} 
+ 2e^{2}\ell^{2} {\bf E}_{\parallel}^{2}/\omega^{2}_{\rm c}}.
\end{equation}
This shows, unexpectedly, that the zero-mode level gap is controllable 
with an injected Hall current.

For a rough estimate let us note the following: 
If we take $\mu \sim 0.05$,  the level gap is $\omega_{\rm c}\mu\,  z 
\sim 5 \times 10^{-3}$meV at   $B= 1$T, with 
$\mu z \sim  1.3\times 10^{-3} B[{\rm T}]$; 
an applied field of strength $E_{\parallel} \sim 1$V/cm leads to a level gap of roughly 
the same magnitude, as seen from 
\begin{equation}
\sqrt{2} e \ell |{\bf E}|/\omega_{\rm c} \sim 10^{-3} \times E[{\rm V/cm}]/ 
B[{\rm T}]^{3/2}.
\end{equation}

One can calculate the spectral weight $\sim |\langle 0''|\rho_{\bf p} |1''\rangle|^{2}$ 
and derive the susceptibility function associated with 
the $1''\rightarrow 0''_{+}$ transition,
\begin{eqnarray}
(\alpha_{\rm e} [{\bf p}])^{0''_{+}}_{1''}&=& {g_{\rm s}e^{2}\over{2 \pi\, \triangle \epsilon}}\,
e^{-x}\, 
{1 + (x/4) \, {\cal E}^{2}/\kappa^{2}\over{1 + {\cal E}^{2}/\kappa^{2}}},  
\end{eqnarray}
with $x= \ell^{2}{\bf p}^{2}/2$ and spin degeneracy $g_{\rm s}=2$.
(Here we have set $\lambda\rightarrow 0$ for clarity; 
there is no appreciable change for $\lambda\sim 0.3$.)
This $\alpha_{\rm e}[{\bf p}]$, with its scale set 
by the tiny gap $\triangle \epsilon \ll \omega_{\rm c}$, 
essentially governs the dielectric property of the $\nu=2$ state.


\begin{figure}[tbp]
\includegraphics[scale=0.9]{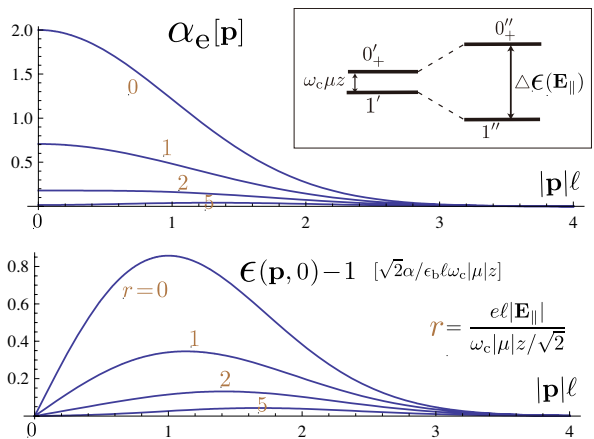}
\caption{ (Color online)  $\alpha_{\rm e} [{\bf p}]$ and 
$\epsilon ({\bf p},0) -1$ at $\nu= 2$, 
in units of  $e^{2}/(2 \pi\,\omega_{\rm c}|\mu|z)$ and 
$\sqrt{2} \alpha /(\epsilon_{\rm b}\ell\, \omega_{\rm c}|\mu| z)$, respectively,
for inplane electric fields of strength 
$\sqrt{2}e \ell|{\bf E}_{\parallel}|/(\omega_{\rm c}|\mu|z) =$ 0,1,2 and 5.
An inset depicts an enhancement of level splitting by an inplane field.
}
\end{figure}


Figure~4 shows $\alpha_{\rm e} [{\bf p}]$ and 
the associated dielectric function
$\epsilon ({\bf p},0) -1$ at $\nu=\nu_{1''} = 2$.
It is seen that both of them decrease rapidly 
as ${\bf E}_{\parallel}$ becomes strong
in the sequence $\sqrt{2}e \ell|{\bf E}_{\parallel}|/(\omega_{\rm c}|\mu|z) 
=0, 1, 2, 5$; this is because the $1''$ and $0''_{+}$ modes are chosen
so as to diagonalize $\int d^{2}{\bf x}\,A_{0}\rho$ (for $\kappa \rightarrow 0$).
In Fig.~4 both $\alpha_{\rm e}[{\bf p}]$ and $\epsilon ({\bf p},0) -1$ 
are plotted in units $1/ |\mu z| \sim (40/|\mu|)(1/B[{\rm T}])$ times as large as 
those in Fig.~2.
Numerically, for $\mu\sim 0.05$ the level gap $\sim 60$ mK at $B=1$T 
would become as large as $\sim$ 300 mK ($\sim$ 6.2\, GHz $h$) by an inplane field of $E_{\parallel} \sim 5$ V/cm. 
This would lead to $\alpha_{\rm e} [0] \sim 14$ and (the peak value of 
$\epsilon ({\bf p},0) -1) \sim 30$ in common units, 
i.e., one-order of magnitude larger than those in Fig.~2.
This suggests that the dielectric effect would show a marked enhancement
around $\nu\sim 2$ if the level gap is resolved by an inplane field
at very low temperatures.

Actually, a larger gap, which leads to better level resolution, 
tends to suppress the effect.
Alternatively it will be more practical to observe the field-induced level gap
$\triangle \epsilon ({\bf E}_{\parallel})$
via the QH effect with an injected current;
one would be able to resolve the $\nu=\pm 2$ Hall plateaus 
using a suitably strong current 
(that may even suppress the dielectric effect).
It will also be possible to detect the level gap directly 
by micro-wave absorption or via conductance modulations~\cite{DCNNG} associated with it.

\section{Summary and discussion}

In this paper we have studied the electromagnetic response of 
bilayer graphene in a magnetic field at integer filling factor $\nu$, 
with emphasis on clarifying the similarities and differences in quantum features 
between monolayers and bilayers.
Bilayer graphene has a unique feature that its band gap is externally controllable;
this makes bilayers richer in electronic properties.

The particle-hole picture of the vacuum state is one of the basic features 
specific to graphene, both monolayers and multilayers, and is not shared 
with conventional QH systems.
In graphene even the vacuum state responds to an external field and acts as a 
dielectric medium, with the Coulomb interaction being efficiently screened 
over the scale of the magnetic length.
Graphene bilayers and monolayers are qualitatively quite similar  
in the vacuum dielectric characteristics (apart from some differences 
that reflect  the underlying Landau-level structures), as we have seen 
in Sec.~III, but numerically the dielectric effect (for the vacuum and 
for $\nu \not=0$ as well) is generally much more sizable for bilayers 
because of the difference in the basic Landau gap,
$\omega_{\rm c}^{\rm bi}/\omega_{\rm c}^{\rm mono} \approx
0.1 \sqrt{B[{\rm T}]}$.

The presence of the zero-energy Landau levels is another basic feature 
specific to graphene.
The monolayer supports four zero-mode levels (one per valley and spin),
while the bilayer supports eight such levels 
(two per valley and spin).  
The zero-mode levels carry normal Hall conductance $e^{2}/h$ per level,
but in monolayers  their effect is hardly visible in density response.
In contrast, in bilayer graphene a gate-controlled interlayer field acts 
to open a band gap 
between the zero-mode levels at the two valleys and this valley gap, in a sense,
activates them:
The dielectric effect due to the pseudo-zero-mode levels grows linearly 
with the band gap and becomes dominant for large gaps, 
as we have seen in Sec.~III.

A finite band gap introduces an asymmetry in the zero-mode spectrum
while it leaves other levels practically symmetric. 
The two zero-mode levels (per spin) at each valley thereby remain degenerate,
apart from a tiny kinetic splitting.  
This robustness in the zero-mode degeneracy at each valley
as well as the presence of the zero modes itself is
a consequence of the nonzero index~(\ref{indexth}) 
of the basic bilayer Hamiltonian
(with $A_{0}\rightarrow 0$ and $z\rightarrow 0$). 
In Sec.~V we have pointed out that this tiny zero-mode level gap at each valley
is enhanced (and even controlled) by an inplane electric field 
or by an injected current.
Such a gap, if properly enhanced by an injected current, may be detected  
via the QH effect or directly via micro-wave absorption.

Finally, in Sec.~IV we have noted that the low-energy characteristics 
of bilayer graphene are neatly summarized by
an effective Chern-Simons gauge theory, which 
accommodates graphene and standard QH systems equally well.

\acknowledgments

The authors wish to thank A. Sawada for useful discussions, especially on 
detection of field-induced level splitting. 
This work was supported in part by a Grant-in-Aid for Scientific Research
from the Ministry of Education, Science, Sports and Culture of Japan 
(Grant No. 17540253).


\end{document}